# Tunnel Barrier to Spin Filter: Electronic Transport Characteristics of Transition Metal Atom Encapsulated in Smallest Cadmium Telluride Cage

Kashinath T Chavan[a,b], Sharat Chandra[a,b] and Anjali Kshirsagar[c]

[a]*Materials Science Group, Indira Gandhi Centre for Atomic Research, Kalpakkam 603102, Tamil Nadu, India.*
[b]*Homi Bhabha National Institute, Training School Complex, Anushakti Nagar, Mumbai, 400094, India*
[c]*Department of Physics, Savitribai Phule Pune University, Pune 411 007, India.*

## Abstract

We report first principles theory-based comparative electronic transport studies performed for an atomic chain of Au, bare $Cd_9Te_9$ cage-like cluster and single transition metal (TM) (Ti, V, Cr, Mn, Fe, Co, Ni, Cu, Zn, Ru, Rh, Pd) atom encapsulated within the $Cd_9Te_9$ using Au(111) as electrodes. The bare cluster is semiconducting and acts as a tunnel barrier up to a particular applied bias and beyond that the device has linear current-voltage relationship. Several TM (Ti, V, Cr, Mn, Fe) encapsulated in the cage show half-metallic behavior and spin filtering effect in the I-V characteristics of the device. A detailed qualitative and quantitative analysis of I-V characteristics for metallic, semiconducting, and half-metallic nanostructures has been carried out.

## Introduction

The field of molecular electronics has received much attention from researchers since the idea to use molecules in the electronic circuit was proposed in 1974 [1]. Various different molecules have been proposed for several spintronics applications, such as switching [2], spin filtering [3-4], GMR [5], DNA sequencing [6], single molecular magnet [7] etc. Remarkably, the organic molecules are being studied extensively due to their longer coherent spin transportability [8]. Along with the molecules, other nanostructural forms such as atomic clusters are also of interest for these applications. Therefore, in addition to molecules, such atomic clusters have also been employed for spintronics applications [5, 9-12] and it is found that the clusters also can show behavior similar to the molecules. However, unlike molecules, the bonding in clusters can be of different types (viz., Van der Waals, ionic, covalent, metallic); they do not have a well-defined geometry; they do exhibit numerous isomers with unique properties that are sensitive to size and shape; and hence transport in such systems can be of fundamental interest apart from applications. After discovering the fullerene cluster, there has been a constant search for similar cage-like structures due to technological importance. The electronic structure and stability of different semiconducting and metallic cage-like clusters have been reported in the literature [13-16].

By encapsulating magnetic atoms such cage-like structures can show unusual behavior. In most cases, such clusters act as molecular magnets, but in some cases, the magnetic moment on the magnetic atom may also get quenched [10]. Encapsulation of metal atoms in clusters has also been reported to help in increasing its stability [13, 16]. Studies of spin-polarized electronic structure of such doped clusters are essential to understand the underlying physics and find their applications in areas like spintronics. The generation and detection of spin by electrical means is the core agenda of research in the field of spintronics. Spin filters, which allow electrons with only one spin to pass and block the other spin, are crucial and are useful devices. Small clusters of CdTe beyond a specific size stabilize in cage-like structures [17], and have enough space for encapsulation. Therefore, appropriate metal atom(s) can be encapsulated in these cages to tune the electronic structure.

Cadmium telluride is a II-VI compound semiconductor with a zinc blend structure and has a direct bandgap of ~1.5 eV [18]. It has applications mainly in solar cells, dilute magnetic semiconductors, and radiation detectors. CdTe is a material that possesses an amphoteric nature that can be tuned as p-type and n-type [19]. Transition metal (TM) doping in CdTe is reported to produce significant polarization in the density of electronic states (DOS) [20-22], and therefore it can be a potential candidate for spintronics applications. TM atom(s) doping in $Cd_9Se_9$ cage structure has been reported to result in spin polarized electronic structure and such doped cages can be stacked to produce more stable structures [23].

This report presents the spin resolved electronic transport characteristics of some 3$d$ and 4$d$ TM encapsulated within cage-like $Cd_9Te_9$ cluster using the Au(111) planes as electrodes. For reference, transport characteristics of the bare $Cd_9Te_9$ cluster and Au atomic chain are also studied and presented. The observed spin filter effect for the $Cd_9Te_9$ cage encapsulating TM atom(s) has been discussed.

**Methodology**
The geometry of the $Cd_9Te_9$ cluster is obtained using the method of simulated annealing. These simulations are carried out by performing ab-initio molecular dynamics (AIMD) using VASP code [24-25]. The detailed procedure used to obtain a possible ground state geometry of the cluster is discussed in Chavan et al. [22] and such TM atom doped $Cd_9Te_9$ cage is used for further transport studies. The electrodes used for this purpose are three layers of Au (111) which are derived from an optimized primitive cell. To form the point contact between the electrode and cluster, a pyramidal cluster of Au is used, which forms the apex of the electrode. The tip of the Au pyramidal cluster makes a point contact with the surface Te atom of the $Cd_9Te_9$ cluster. This device is divided into three parts, viz., two electrodes (left and right) and a scattering region or central region. The scattering region has a cluster of CdTe sandwiched between the two Au pyramidal clusters (electrode extensions). For the transport characteristics of the Au atomic chain, the cluster is replaced by an Au atomic chain of similar size with initial interatomic separation corresponding to the bulk (2.94 Å) [26].

The calculations of device geometry optimization, transmission spectrum, projected device density of states and I-V characteristics, etc. are carried out using first-principles density functional theory coupled with non-equilibrium Green's function formalism as implemented in the Quantum-wise ATK package [27-29]. To perform these calculations, FHI pseudo-potentials are used [30]. Linear combination of atomic orbitals (LCAO) with double zeta polarized (DZP) basis set is employed for this purpose. Spin-polarized calculations are performed within GGA-PBE to approximate the exchange-correlation energy functional [31-32]. All the device geometries are relaxed to the maximum force tolerance of 0.02 while constraining the 3 outer layers of the electrodes to be rigid. Since there are elements with complex *d* shell electrons viz., TM atoms, these calculations have been performed using quite fine density mesh with mesh cutoff of 300 Ha. The criterion used for achieving self-consistency is $10^{-4}$ eV. The k-mesh grid of 1×1×171 points is used for geometry optimization and a finer transverse grid of 9×9 points for transmission spectrum calculations. A set of TM (Ti, V, Cr, Mn, Fe, Co, Ni, Cu, Zn, Ru, Rh, Pd) atoms are separately encapsulated or endohedrally doped in the cluster.

The transmission spectrum is calculated using the following formula,

$$T(E, V) = Tr[\Gamma_L(E, V)G^R(E, V)\Gamma_R(E, V)G^A(E, V)]$$

where $\Gamma_L$ and $\Gamma_R$ are the coupling matrices between the left/right electrodes and the central region. $G^R$ and $G^A$ are retarded and advanced Green's function in the scattering region. T(E, V) represents the transmission probability of an electron with energy E to be transported from one electrode to the other with bias voltage V. The current-voltage characteristics is calculated as per the following formula,

$$I(V) = \frac{2e}{h} \int T(E, V)[f(E - \mu_L) - f(E - \mu_R)]dE$$

where *e* and *h* are electronic charge and Planck's constant respectively, *f*(E) is the Fermi function and $\mu_{L/R}$ is the chemical potential of left/right electrode. For spin-polarized calculations, quantities T(E, V), $\Gamma_L$, $\Gamma_R$, $G^R$, $G^A$ and I(V) have two components corresponding to up and down spins of electrons.

**Results and Discussion**
We report the electronic transport studies of three different (electronic structure wise) classes of materials in a nanostructured form, viz., metallic (Au atomic chain), semiconducting (atomic cluster of $Cd_9Te_9$), and half-metallic (single TM encapsulated in cage-like cluster of $Cd_9Te_9$) using the Au(111) electrodes. This transport study calculates the spin-resolved projected device density of states, transmission spectrum, and I-V characteristics for these devices. The transport properties are primarily dependent on transmission around the Fermi energy ($E_F$) especially for low biases.

The atomic chains of monovalent Au are reported to be stable and have quantized conductance which depends on the number of constituting atoms and stress in the chain [33].

In the present study, the number of atoms constituting the chain is chosen such that its length is equal to the diameter of the cluster and therefore the scattering central region is of almost the same width in the three cases considered for the study. Figure 1 shows the model device with Au atomic chain and Au (111) electrodes.

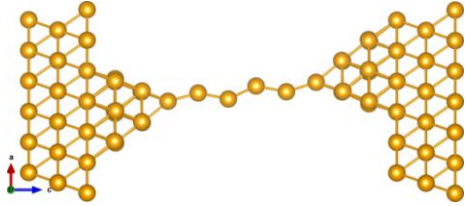

Figure 1.The model device of Au atomic chain and Au(111) electrode. The apex of the pyramidal Au cluster connects the electrodes to the chain, the distance between the tips of the electrodes being 13.19 Å.

It is observed that the initial simple linear chain acquires zig-zag form upon relaxation of the device, which is the stable form of the Au atomic chain with two nearest neighbors over the simple linear one [34]. The interatomic distance in the chain after relaxation of the device becomes 2.69 Å. These atomic separations in the chain are consistent with the earlier reported value [26] (2.62 Å) within an error ~2%. This atomic chain is treated as a nanostructure-like molecule and its eigenstates corresponding to HOMO-1, HOMO, LUMO and LUMO+1 energies are plotted in Fig. 2. The occupied states are more localized, and the orbital features are clearly seen; in comparison the unoccupied states are hybridized.

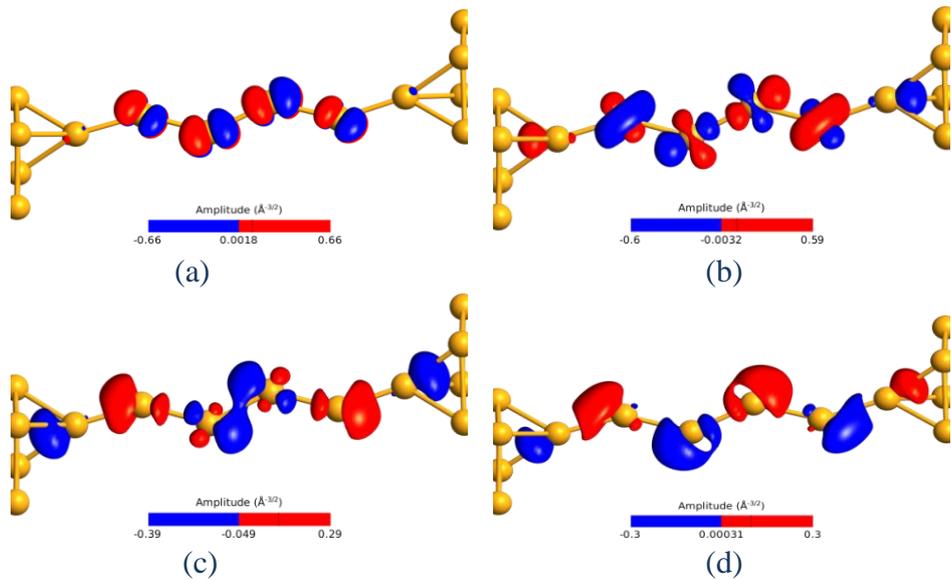

Figure 2: The eigenstates corresponding to (a) HOMO-1, (b) HOMO, (c) LUMO and (d) LUMO+1 energies for Au atomic chain. [iso value: 0.1]

The transmission spectrum and I-V characteristics for Au atomic chain are shown in Fig. 3. Au chain has non-zero transmission in the energy window of -2 eV to +2 eV about $E_F$. This transmission spectrum is qualitatively consistent with earlier reported spectrum [35]. The I-V characteristic is calculated for the bias range of 0 V to 1 V in steps of 0.1 V. As expected, the I-V characteristic for Au atomic chain is linear.

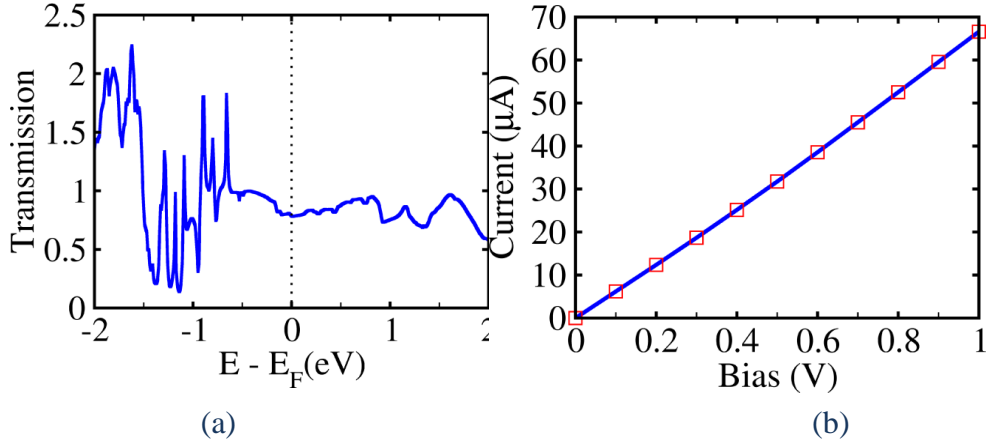

Figure 3. (a) Transmission spectrum and (b) I-V characteristics for the device with Au atomic chain. Transmission is calculated for the energy window of -2 to +2 eV about the Fermi energy $E_F$.

The second type of nanodevice consists of the cage-like semiconducting cluster $Cd_9Te_9$ and is shown in Fig. 4. This CdTe cluster exhibits $C_3$ symmetry and its outermost atoms are Te. Therefore, the point contact is made between electrode extension (apex of pyramidal Au cluster) and the outer Te atoms. The length of this central region (Au-Au) is 13.42 Å. The Au-Te bond length at the tip is 2.80 Å.

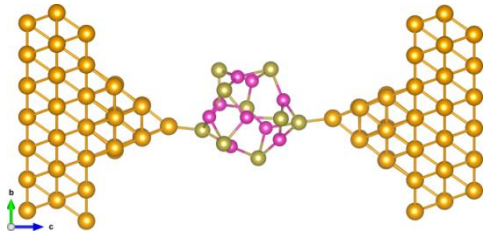

Figure 4. The model device with the $Cd_9Te_9$ cluster and Au(111) electrodes. The contact atoms are Te (olive green).

Figure 5 shows the eigenstates corresponding to the HOMO-1, HOMO, LUMO and LUMO+1 energies of the pristine $Cd_9Te_9$ cluster. These states are mostly populated by Te $p$ states; however, the unoccupied states are hybridized.

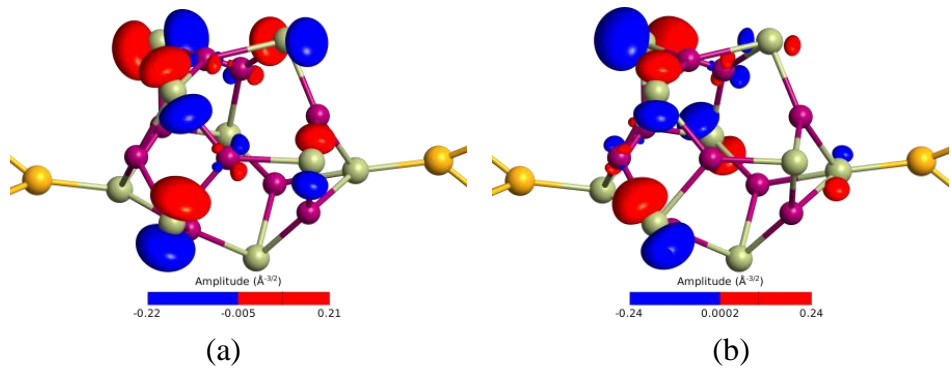

(a)    (b)

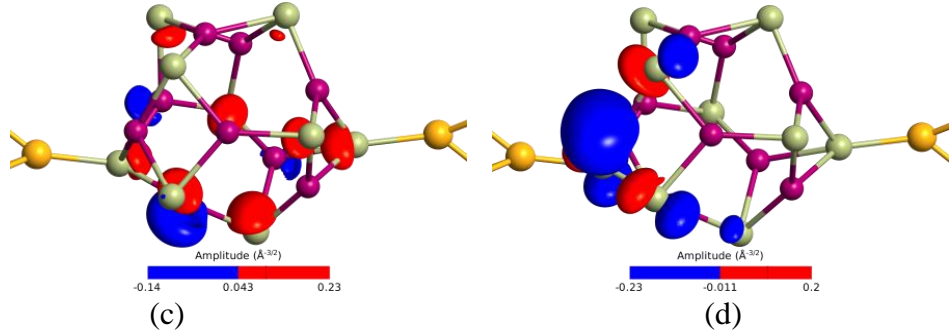

Figure 5. Eigenstates corresponding to (a) HOMO-1, (b) HOMO, (c) LUMO and (d) LUMO+1 energies of sandwiched $Cd_9Te_9$ cluster [iso value: 0.1]. The results are identical for up and down spins, as expected. The Cd atoms are in magenta, Te in olive green and Au in yellow.

Figure 6 shows the transmission spectrum and I-V characteristics of semiconducting $Cd_9Te_9$ cluster with Au electrodes. The transmission is zero in the energy window of ~1.6 eV about $E_F$ (corresponding to the energy gap of the cluster). Therefore, the small current in I-V characteristics for a bias range of 0 V to 1.6 V is the tunneling current. For higher bias voltages, the charge carriers overcome the energy gap of the cluster and the current shoots up in the device as the contact becomes ohmic. Therefore, there is a sudden increase in current beyond 1.6 V. Thus, the semiconducting cluster sandwiched between the Au electrodes acts as a tunnel barrier for the low bias range.

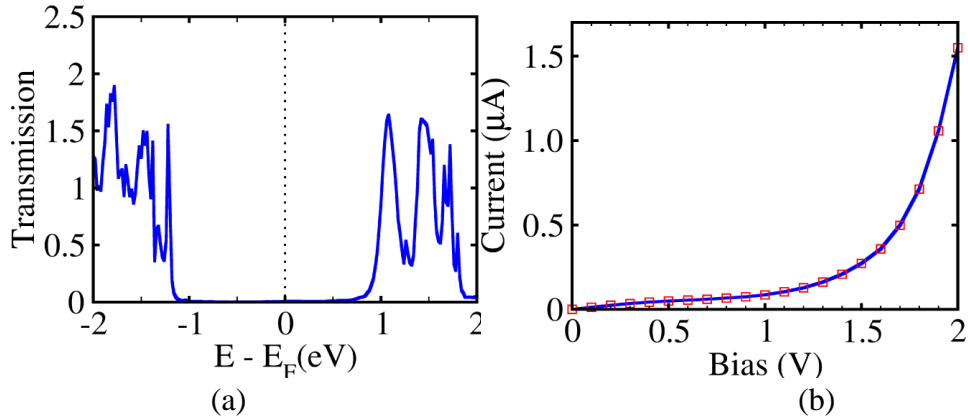

Figure 6. (a) The device's transmission spectrum and (b) I-V characteristic of semiconducting $Cd_9Te_9$ cluster as a barrier and Au(111) as electrodes. The transmission spectrum is calculated for the energy range of -2 eV to +2 eV about $E_F$ and the I-V characteristic is calculated for a range of bias voltages from 0 V to 2 V in steps of 0.1 V.

Next, we report the interesting study of the electronic transport properties of endohedrally TM atom(s) doped $Cd_9Te_9$ cage-like cluster. Such a model device of semiconducting cage $Cd_9Te_9$ with a TM atom encapsulated inside and connected to the Au electrodes is shown in Fig. 7. For encapsulation, atoms with less than half-filled, exactly half-filled, and more than half-filled outer $d$ states from $3d$ elements (Ti, V, Cr, Mn, Fe, Co, Ni, Cu, and Zn) are employed. The $4d$ elements have more de-localized outer $d$ states than the $3d$ elements. In order to understand its effect on electronic transport, a few $4d$ atoms (Ru, Rh, and Pd) have been used for encapsulation inside the $Cd_9Te_9$ cage cluster in the device.

In TM doped cage, the Cd-Te bond length ranges from 2.78 Å to 3.17 Å, whereas without encapsulation, it is 2.77 Å to 2.95 Å. At the tip of the pyramidal Au cluster, Au-Te bond length remains close to the average value, i.e., 2.72 Å for all cases. The distance between the two tips (Au-Au) of electrode extension (pyramidal Au cluster) is close to 13.00 Å. The encapsulating cluster geometry is largely distorted for Fe, Co, and Ni atoms among the 3$d$ elements and for Ru atom among the 4$d$ elements. For the remaining cases, the encapsulated TM atom(s) remain at the geometric center of the cage; albeit the cage structure is slightly changed. The Fe, Co, Ni and Ru atoms have moved away from the cage-center but still are inside the cage.

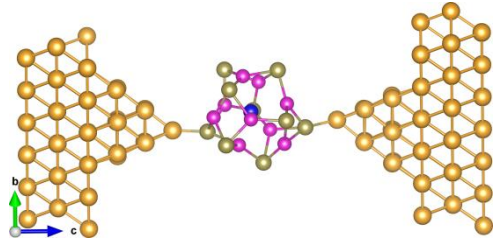

Figure 7. The model device of TM atom encapsulated (blue) inside Cd$_9$Te$_9$ cage-like cluster with Au(111) electrodes. The contact atoms are Te.

Spin resolved projected device density of states (PDDOS) of TM doped Cd$_9$Te$_9$ cage in the device is shown in Fig. 8. For reference, PDDOS for the bare cluster is shown at the bottom for each panel. The orbital resolved density of states (DOS) is given in Fig. S1 of supplementary information (SI). In the given energy range, the contribution to DOS due to Cd is negligible compared to Te and TM. The states around E$_F$ are mainly populated by TM states whereas Te $p$ states start dominating as one moves away from E$_F$. The encapsulated TM atom $d$ states predominantly lie around the Fermi energy except for the completely filled $d$ states atoms Cu, Zn, and Pd as well as for Rh atom. TM $d$ states lie below the Fermi energy for these systems. In these systems, the PDDOS is almost same for up and down spins and therefore these systems are not of interest for spin-based devices. The non-zero states at the Fermi energy are predominantly due to Te $p$ and Pd $s$ states for Pd for both the spins; therefore, Pd encapsulation has changed the nature of the cluster from semiconducting to metallic. Significant polarization in DOS is observed in most of the other TM encapsulated in Cd$_9$Te$_9$ cage.

For V, Mn and Ru, $d$ states due to only one spin are present at the Fermi level (PDDOS due to the other spin is zero) suggesting that the nature of cluster in the device is half-metallic. Ti, Cr, Fe, and Co have significantly different PDDOS in the vicinity of the Fermi level for the two spins. The polarization in PDDOS can be quantified in terms of magnetic moment μ$_{TM}$ on encapsulated TM atoms. μ$_{TM}$ in the device increases from Ti to Mn and then decreases as expected. The highest magnetic moment is for Mn (3.8 μ$_B$) followed by Cr (3.7 μ$_B$), Fe (2.7μ$_B$), V (2.6 μ$_B$), Co (1.8 μ$_B$), Ti (1.5 μ$_B$), Ni (0.54 μ$_B$), Ru (0.56 μ$_B$), etc. These magnetic moment values are consistent with the Slater-Pauling curve of magnetization on TM in the nanostructures. Mn atom has exactly half-filled $d$ orbital and Cr atom has exactly half-filled $d$ and $s$ orbital as per Hund's rule; these systems therefore have the maximum magnetic moment. The decrement in the magnetic moment on the neighboring elements is almost symmetric considering the pairs (V, Fe) and (Ti, Co). Co and Ru atoms have 7 electrons in their *d* states and Ru further has a single electron in 5*s* states but quenching is more for the 4d TM atoms. There is no magnetic moment on the atoms Cu, Zn, Rh, and Pd as expected from the PDDOS. Overall, the magnetic moment is systematically quenched for all the atoms in the device.

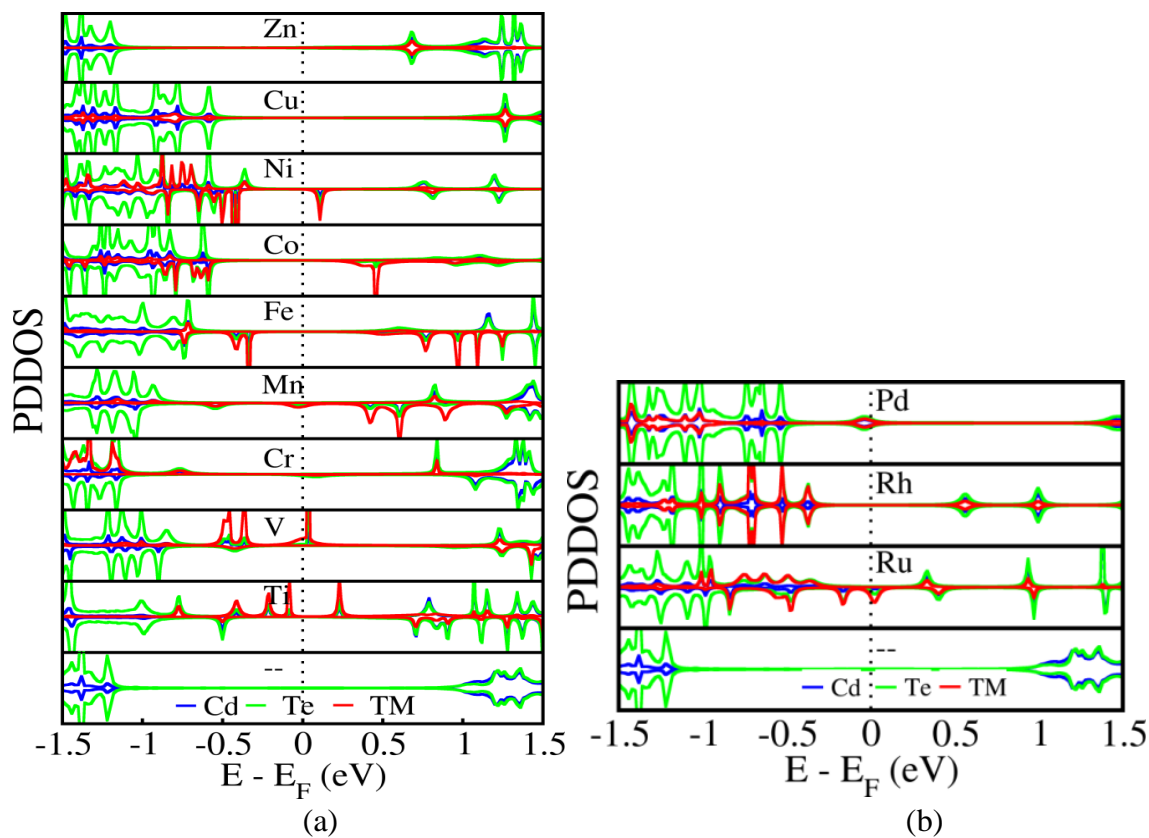

Figure 8. Spin resolved, atom projected device density of states (PDDOS) for (a) 3d (Ti, V, Cr, Mn, Fe, Co, Ni, Cu, Zn) and (b) 4d (Ru, Rh, Pd) TM atom(s) encapsulated in cage-like $Cd_9Te_9$ cluster with Au electrodes. The atom projected device DOS is shown for Cd (blue), Te (green), and the encapsulated TM atom (red), as indicated at the bottom of the figure. PDDOS axis range is [-20:20] states/eV. The curves above (below) the energy axis represent up (down) spin states.

The transmission spectra of the device, as shown in Fig. 9 for all the doped systems, are mostly consistent with the PDDOS plots. Polarized transmission channels for spin carriers exist at the Fermi level in only V, Cr, Mn and Ru doped cage. However, V doped cage has a peculiar behavior, namely the PDDOS is non-zero at $E_F$ for up spin but the transmission spectrum has channels for down spin near $E_F$. Atomic Mn states have the energy ordering as 4*s* up, 3*d* up, 3*d* down and 4*s* down and accordingly there are no states available near $E_F$ for up spin in PDDOS. However, the transmission is non-zero just above $E_F$ for up spin. PDDOS for Cr doped cage device is small for both spins near $E_F$ but the transmission is more for down spin. Non-zero transmission channels are present for one spin carrier in close vicinity of $E_F$ and a quite large gap for the other spin channel in the Ti, Fe, Co and Ni doped cage.

PDDOS and transmission spectra convey same information for Ni, Cu, Zn, Rh and Pd doped cage.

In addition to transmission spectrum, the transmission eigenstates at the Fermi level for each spin are plotted for the V, Cr, Mn and Pd in Fig. 10 (a) to (h) and for Ti, Fe, Co, Ni, Cu, Zn, Ru, Rh in Fig. S2 (a) to (p) in SI. The corresponding transmission eigenvalues are tabulated in Table ST1 in SI. In both the V and Ti doped cases, only the majority spin states are present at $E_F$ with almost no states due to minority spin carriers. These arise from the V $d$ and Te $p$ electrons in V doped cage whereas, they arise only from the Ti $d$ electrons in Ti doped cage. For the rest of the cases, transmission states present at $E_F$ get contributions mainly from the minority spin carriers. The down spin states in Mn, Fe, Co and Ni are mainly the TM $d$ states, while for the Cr and Ru, the states exhibit hybridization between the TM and the Te states. The transmission eigenstates at $E_F$ have transmission channels due to the Te $p$ and Pd $s$ states for Pd for both the spins. Hence Pd doped cage is expected to show metallic behavior for both the spins.

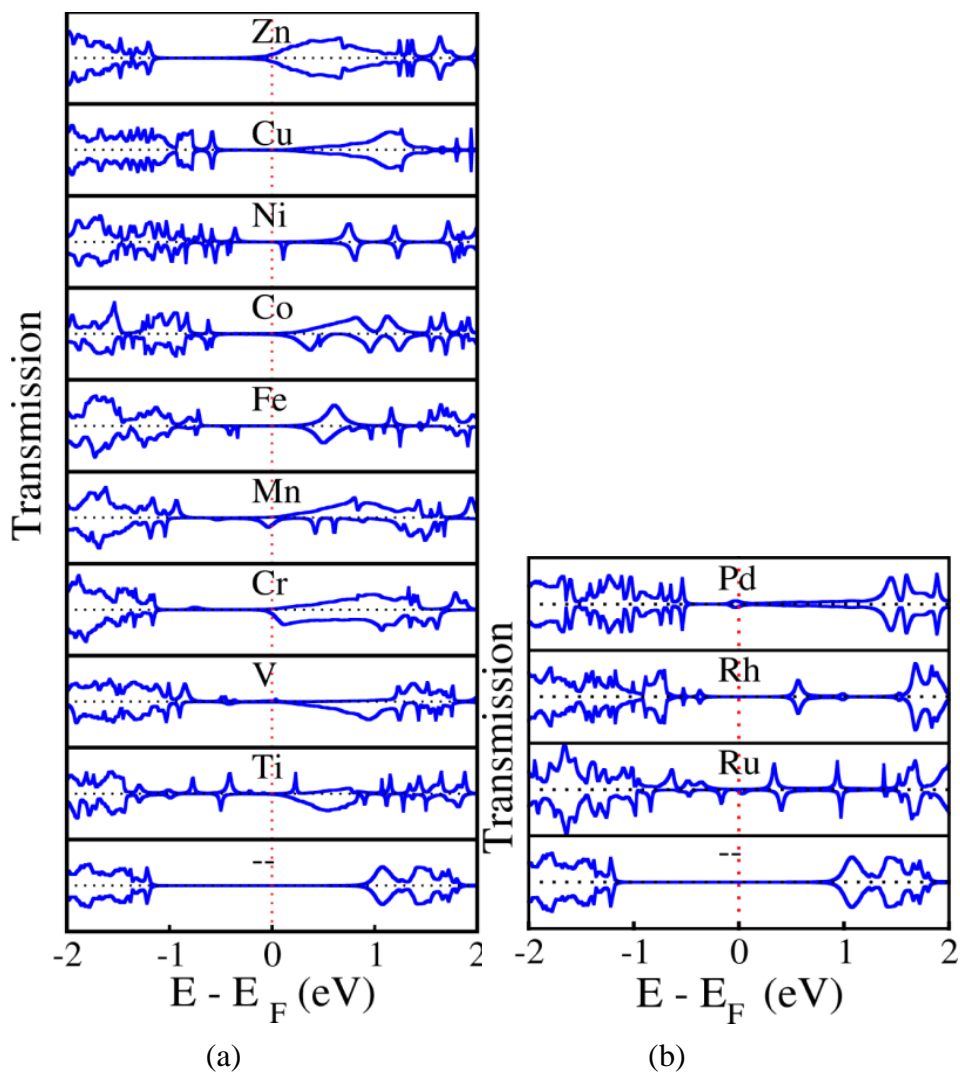

Figure 9. The device's transmission spectrum with TM encapsulated inside $Cd_9Te_9$ cage and Au(111) electrodes. This sandwiched cluster encapsulates (a) 3d (Ti, V, Cr, Mn, Fe, Co, Ni, Cu, Zn) and (b) 4d (Ru, Rh, Pd) TM

atom(s). The encapsulated TM atom is mentioned in the respective transmission spectrum plot. For reference, the transmission spectrum for the bare cluster is shown at the bottom in both the panels.

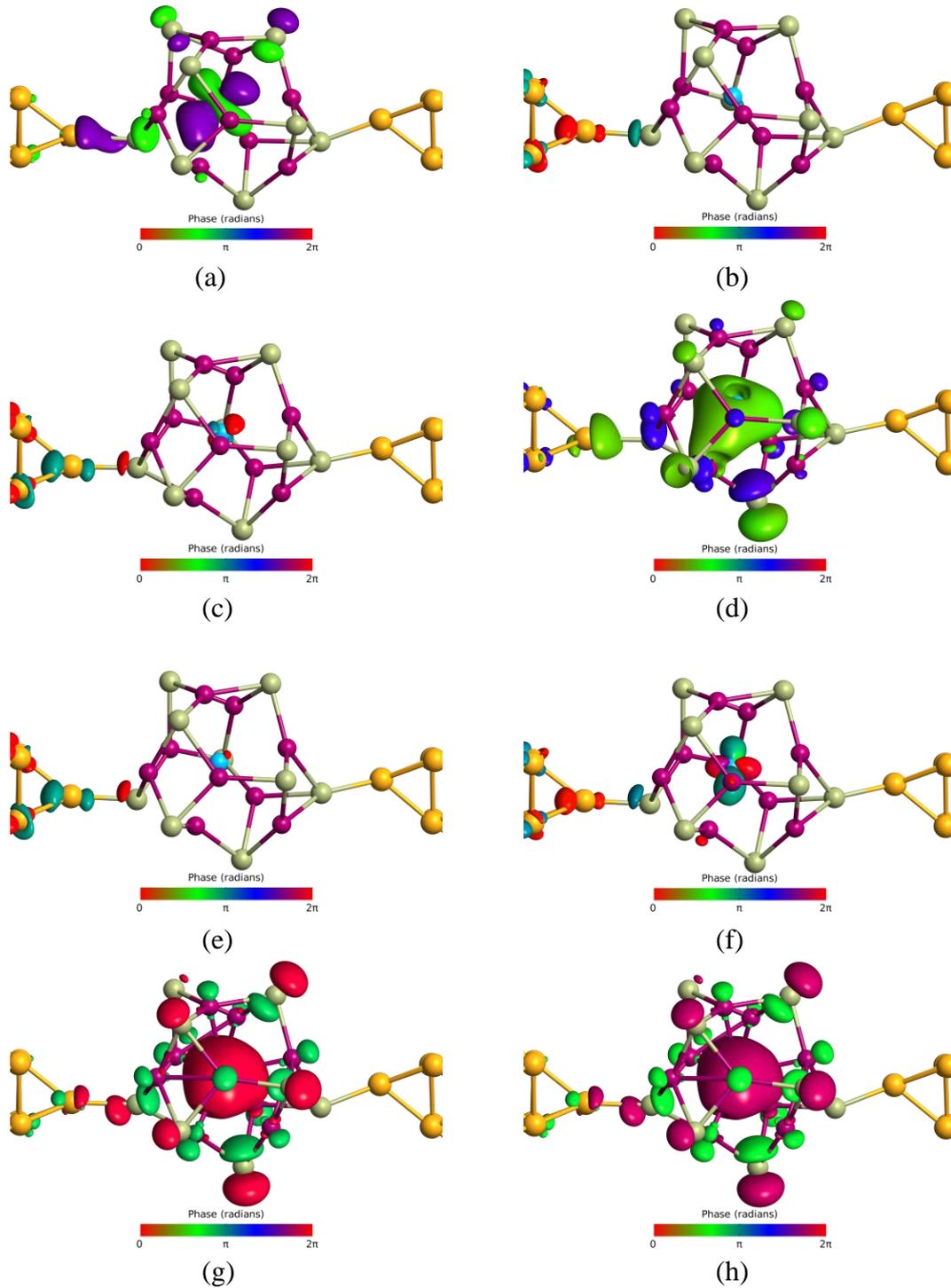

Figure 10: Transmission eigenstates for the device with TM atom encapsulated in $Cd_9Te_9$ cluster for (a) V up spin, (b) V down spin, (c) Cr up spin, (d) Cr down spin, (e) Mn up spin, (f) Mn down spin, (g) Pd up spin and (h) Pd down spin. [iso value: 0.4].

Transmission eigenstates probe the probability of transmission for specific energy E and applied bias V, therefore we have also studied the transmission eigenstates for different bias values for specific cases of interest. We could correlate the probability with the I-V characteristics of the respective devices.

Since there is considerable polarization near the Fermi level in PDDOS and the transmission spectrum for specific TM atoms encapsulated in the cage, the I-V characteristics of the device are depicted for those TM (Ti, V, Cr, Mn, Fe, Co, Ni, and Ru) cases only. Figure 11 shows the spin-resolved I-V characteristics of the device for bias range 0.0 V to 1.0 V. For Cr, Mn, Ni, and Ru atoms encapsulated in the cage, the current is only due to minority spin carriers in the low bias range. For the rest of the elements (Ti, V, Fe, Co), current is small up to 0.4 V for both spins. Low current indicates that the molecular orbitals of the doped cage cannot couple with the states of the Au electrodes and no channel for transport is available. There is a difference of a few μA in the current magnitude due to majority and minority spin carriers. There is 100% spin polarization for Mn and Cr doped cages at zero bias, as is also indicated from the transmission spectra at $E_F$ for zero bias. This is also evident from the transmission eigenstates for up and down spins as depicted in Fig. 10 (c)-(f). The current due to majority spin carriers increases slowly and is almost linear. Thus, these devices are good as spin filter for the complete range of bias values studied. For the Ti based device, although the currents due to both spins are small at low bias values, the minority current rises steeply beyond 0.5 V. In the bias range studied, the current due to minority spin carriers dominates throughout in all the cases of interest. However, for Ru, I-V curves of both spin carriers cross each other twice, therefore in the bias range of 0.4V - 0.6V, the current due to majority spin carriers is higher than that of minority spin carriers. In addition, the Ru doped cage also shows the interesting feature of small negative differential resistance (NDR) in the bias range 0.8 V to 0.95 V. NDR is seen only for the Ru doped cage. For the Pd doped cage, currents for both the spins (not shown here) are linear and are equal and this is consistent with the PDDOS, transmission spectra and eigenstates plots (Fig. 10 (g)-(h)).

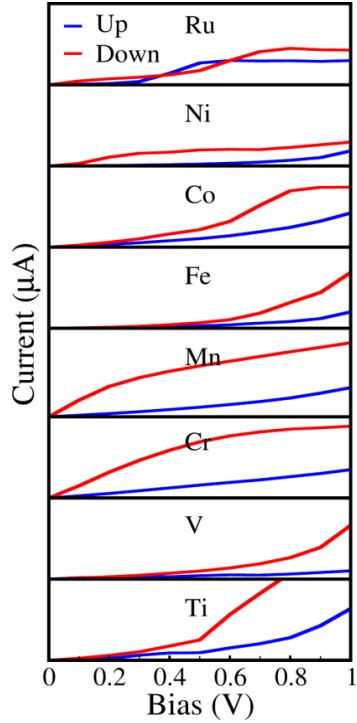

Figure 11. Spin resolved I-V characteristics of selected TM (Ti, V, Cr, Mn, Fe, Co, Ni, Ru) encapsulated inside $Cd_9Te_9$ cage sandwiched between Au electrodes. The I-V characteristics are calculated for the bias range of 0 V to 1 V in steps of 0.1 V. The current axis range is 0 µA to 6 µA for each plot.

In this work, we have studied the electronic transport phenomenon in different nanostructures (metal, semiconductor and half metal) in the device with Au electrodes. The linear chain of Au behaves as if metallic, which is expected, and the results like Au-Au bond lengths, transmission spectra are consistent with the earlier reports [26,35]. The I-V curve is linear and has a maximum current of nearly 70 µA for a bias of 1.0 Volts. The I-V characteristics of this system can be used as a reference for quantitative analysis for the device with semiconducting and half-metallic clusters. The length of the chain was chosen to match the diameter of the $Cd_9Te_9$ cage closely.

Further, the cage-like semiconducting cluster, $Cd_9Te_9$, is employed to study electronic transport properties through transmission spectrum and I-V characteristics. In the transmission spectrum, there are no channels in the energy range of 1.5 eV approximately, about the Fermi energy, which corresponds to the cluster's HOMO-LUMO gap. The I-V curve shows small current - the tunneling current, in the low bias range. It also suggests the suitability of this cluster in tunnel magnetoresistance (TMR) devices, where an insulator is sandwiched between ferromagnetic electrodes. Beyond a threshold bias voltage (1.6 V), the current increases rapidly, suggesting that the electrons overcome the barrier and the contact becomes ohmic. Therefore, there is a linear I-V characteristic beyond this threshold voltage. Since there is no way to distinguish between the majority and minority carriers, PDDOS, transmission spectra and I-V curves are identical for both spins.

Since there is enough space inside the cage, the cluster is mono-doped endohedrally using all 3$d$ and some 4$d$ TM atoms, and their transport characteristics are studied. In V doped cage, the electronic states are non-zero for up spin and are due to V $d$ states, whereas in Cr doped cage, the corresponding states are hybridized Cr $s$ and Te $p$ states. Further, Mn with exactly half-filled 3$d$ states has non-zero states at $E_F$ predominantly due to minority charge carriers of Mn $d$. Among the more than half-filled 3$d$ TM atoms, only Ni has non-zero states at $E_F$ and they are due to Ni $d$. Thus, out of the 3$d$ TM encapsulated cluster device, an exemplary half-metallic character is seen in PDDOS for V, Cr, Mn, and Ni. Apart from these, Ti, Fe, and Co have significant polarization in PDDOS, there are non-zero states near $E_F$ due to one spin carriers and considerable gap for other spin carriers. Cu has filled $d$ shell with one electron in the $s$ shell; still the PDDOS shows no polarization. This may happen due to hybridization of the Cu and Cd $s$ states. As expected, Zn doped cage has filled 3$d$ and 4$s$ shells; the PDDOS does not have any polarization of electronic states at $E_F$. The 4$d$ elemental cases, Ru, Rh and Pd atom encapsulated in cage, show half-metallic, semiconducting and metallic behavior respectively. In the Ru case, there are non-zero states at $E_F$ due to minority charge carriers from $d$ states, and for the Pd case, there are non-zero states at $E_F$ due to Pd $s$ and Te $p$ for both spin channels.

The orbital characters in the transmission spectrum at the Fermi level are seen better in the transmission eigenstates. Overall, the polarization in atom projected partial DOS shows that the encapsulated atom retains its isolated atom-like features when encapsulated in the $Cd_9Te_9$ cluster. The device with Ti, V, Cr, Mn, Fe, Co, Ni, and Ru atoms encapsulated in cage produces spin-polarized current, which is confirmed from the I-V characteristics of respective devices. Cr, Mn and Ni doped cages have current only due to the minority charge carrier in the low bias range, and this is called spin filtering effect, which is essential for spintronics. Ru doped cage also shows polarized currents but with small and comparable magnitudes. The rest of the elements significantly differ in the current magnitude of two spins at higher bias. Therefore, these devices can be the source of spin-polarized current at higher bias values. The maximum current ~70 μA is observed for metallic nano contact at bias of 1.0 V, ~ 5-9 μA at bias of 1.0 V for half-metallic systems and ~ 0.1 μA and ~ 1.6 μA for the semiconducting system at the bias of 1.0 V and 2.0 V respectively.

**Conclusions**
The electronic transport studies have been carried out for three different classes of nanostructures using the Au as electrodes. These nanostructures include metallic Au atomic chain, semiconducting $Cd_9Te_9$ cages, and TM atom(s) encapsulated inside $Cd_9Te_9$ cage. The atomic linear chain of Au relaxes into a zig-zag chain and has linear I-V characteristics with current ~70 μA for 1.0 V bias. The semiconducting $Cd_9Te_9$ cage with Au nanocontacts produces small tunneling current for low biases, which corresponds to the energy gap of the cage, and beyond a threshold bias, the cage becomes ohmic and results in linear I-V characteristics. The encapsulated TM atoms remain inside the cage upon device relaxation. Ti, V, Cr, Mn, Fe, Co, Ni and Ru encapsulated cage(s) have half-metallic nature, and the corresponding devices produce spin-polarized current. Among these systems, in the low bias, Ti and V based devices have very small current but the current increases only due to down spin carriers for bias greater than 0.5 V approximately, which is a feature of a spin filter. Mn and Cr based devices display small current and linear I-V characteristic for up spin but quite large current for the down spin throughout the range studied. Hence, they can be good spin filters, particularly in bias range of 0 - 0.7 V. Fe and Co based devices have a difference in the magnitude of the currents for the two spins, and thus they can be a source of polarized current for higher bias. Ru doped cage shows a negative differential resistance around 0.8 V and therefore can act as a tunnel junction but is not good for use as a spin-based device. Thus, TM atom doped cage-like structures of inorganic substances can also be used in spin-based devices in line with organic molecules.

It is observed from the PDDOS that the encapsulated TM atoms have retained their isolated atom-like features in the cage, which is revealed from a small hybridization of TM states with those of Cd or Te states. The magnetic moments on the TM atoms in the device are consistent with the Slater-Pauling curve. However, larger distortions in the geometries, like for more than half-filled cases (Fe, Co, and Ni), are reflected in the magnetic moments.

**Acknowledgment**
KTC acknowledges DAE/IGCAR for the research fellowship and AK acknowledges DST Nanomission Council for financial support through a major research project to set up a high-performance computing facility (DST/NM/NS-15/2011(G)).

# Supplementary Information

# Tunnel Barrier to Spin Filter: Electronic Transport Characteristics of TM Encapsulated in smallest Cadmium Telluride Cage

Kashinath T Chavan[a,b], Sharat Chandra[a,b] and Anjali Kshirsagar[c]

[a]Materials Science Group, Indira Gandhi Centre for Atomic Research, Kalpakkam 603102, Tamil Nadu, India.
[b]Homi Bhabha National Institute, Training School Complex, Anushakti Nagar, Mumbai, 400094, India
[c]Department of Physics, Savitribai Phule Pune University, Pune 411 007, India.

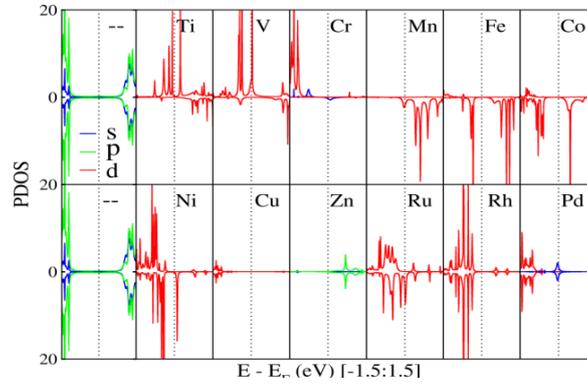

Figure S1. Spin resolved, orbital projected (TM $s$, $p$, $d$) device density of states for TM atom(s) encapsulated in $Cd_9Te_9$ cage sandwiched between Au (111) electrodes. Plots in the 1st column (Cd $s$ and Te $p$) correspond to a bare cage, a reference device. For each plot, energy ranges from -1.5 eV to 1.5 eV about the respective Fermi energy.

| TM | Transmission Eigenvalues (in eV) $\times 10^{-2}$ | |
| --- | --- | --- |
|    | Up spin | Down spin |
| Ti | 0.573 | 3.640 |
| V  | 1.728 | 2.340 |
| Cr | 6.335 | 29.57 |
| Mn | 3.819 | 18.66 |
| Fe | 0.634 | 1.090 |
| Co | 1.368 | 2.326 |
| Ni | 0.205 | 0.256 |
| Cu | 1.560 | 1.560 |
| Zn | 43.24 | 43.24 |
| Ru | 0.742 | 18.00 |
| Rh | 0.192 | 0.193 |
| Pd | 6.165 | 6.180 |

Table ST1: Transmission eigenvalues for device of TM atom(s) encapsulated in $Cd_9Te_9$ cage at Fermi level for up and down spins.

**Ti**

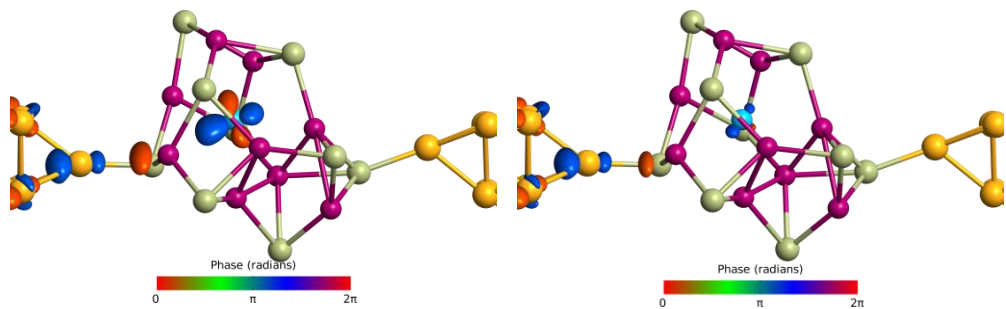

(a)  (b)

**Fe**

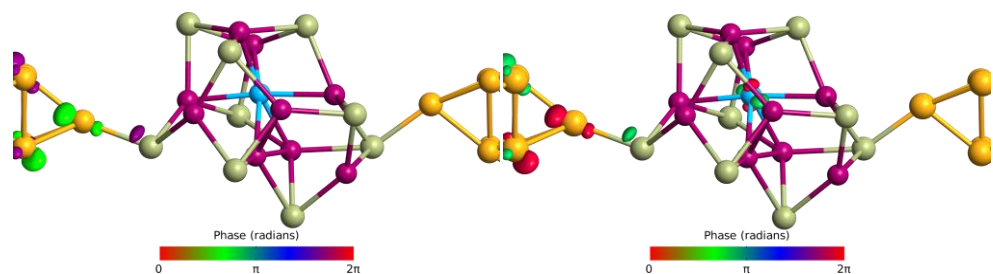

(c)  (d)

**Co**

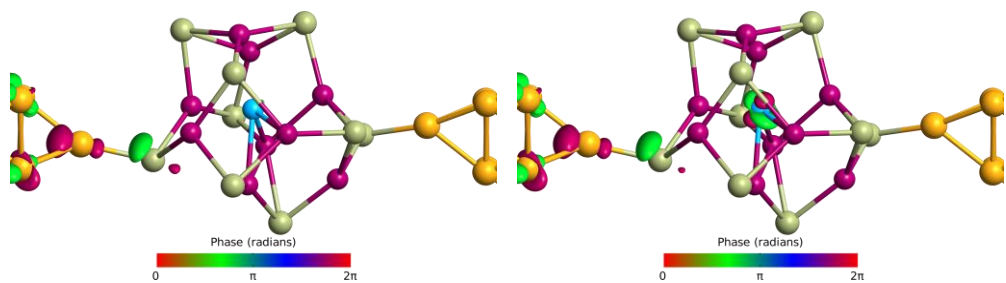

(e)  (f)

**Ni**

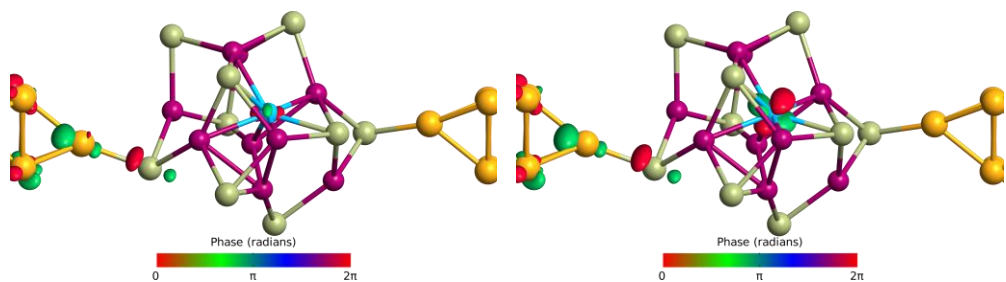

(g)  (h)

**Cu**

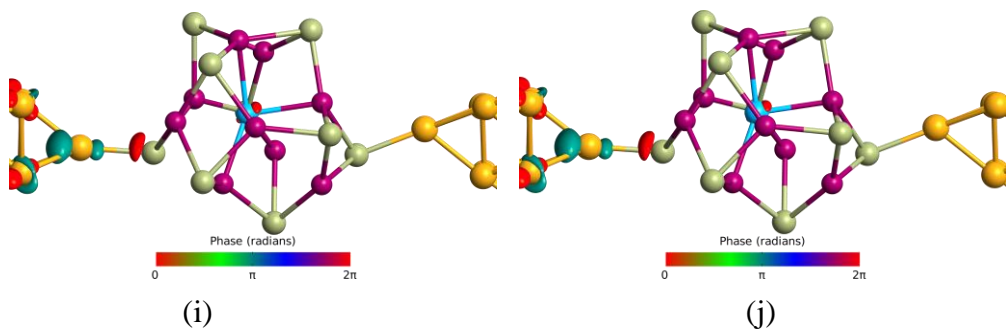

(i) (j)

**Zn**

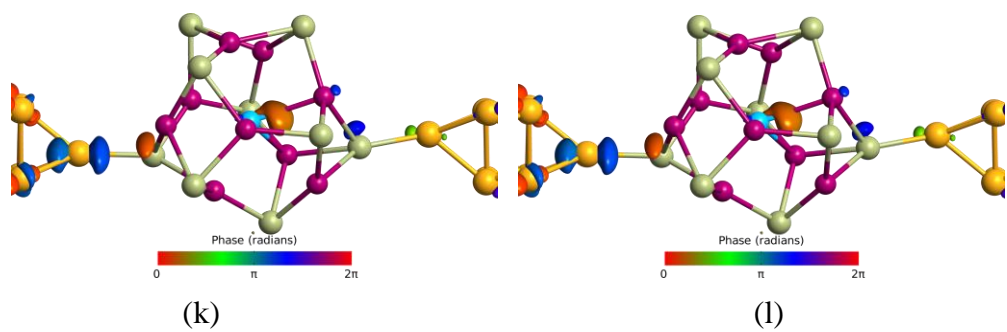

(k) (l)

**Ru**

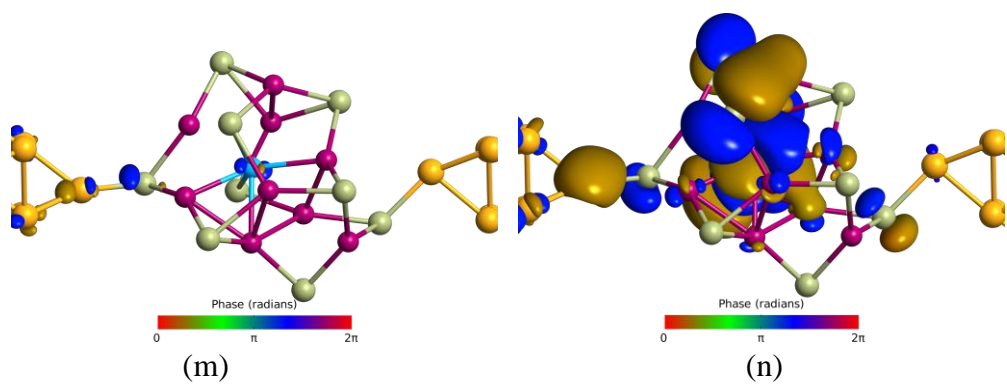

(m) (n)

**Rh**

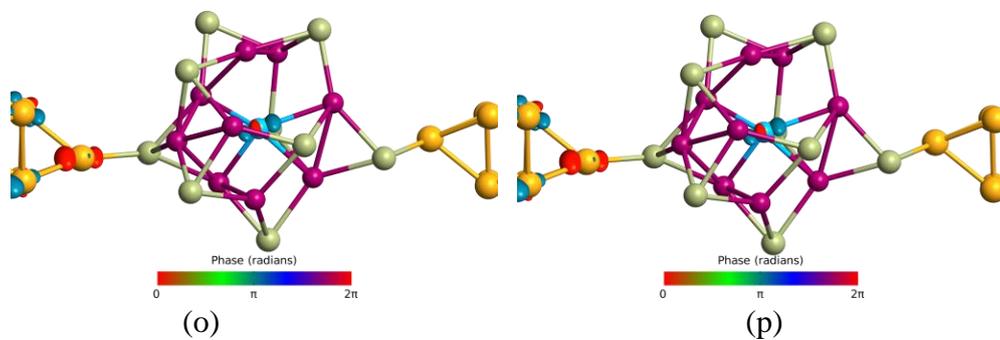

(o) (p)

Figure S2: Transmission eigenstates for the device with respective TM atom encapsulated in $Cd_9Te_9$ cage at Fermi level for (a) up and (b) down spins. (iso value: 0.4).